\documentclass[aps,onecolumn,10pt]{revtex4}
\usepackage{amsmath}
\usepackage{amssymb}
\usepackage{hyperref}
\hypersetup{colorlinks=true,linkcolor=red,citecolor=green,urlcolor=blue}
\numberwithin{equation}{section}
\numberwithin{equation}{section}

\begin{document}
\allowdisplaybreaks
\setcounter{equation}{0}
\baselineskip=0.88cm
 
\title{Exact Solution to Standard Model Hydrodynamic Cosmological Perturbation Theory and its Implications for Acoustic Oscillations}

\author{Philip D. Mannheim}
\affiliation{Department of Physics, University of Connecticut, Storrs, CT 06269, USA \\
philip.mannheim@uconn.edu\\ }

\date{February 17 2025}

\begin{abstract}
We present an exact solution to standard model  cosmological perturbation theory in a matter-dominated, adiabatic, hydrodynamic era. The solution is in the form of  hypergeometric functions. While such functions can oscillate with the sound velocity, they can only do so at high frequency. There is thus a maximum wavelength  to these oscillations, with this maximum wavelength serving to provide a horizon for acoustic oscillations.

\end{abstract}

\maketitle

\section{Introduction}
\label{S1}
Central to cosmological research is the study of  fluctuations around a Robertson-Walker background (see e.g.  \cite{Dodelson2003,Mukhanov2005,Weinberg2008,Lyth2009,Ellis2012}). In this research the focus has been on the Einstein-gravity-based inflationary  de Sitter universe model \cite{Guth1981} (for other relevant de Sitter studies see  \cite{Brout1978,Starobinsky1979,Kazanas1980}), leading to a concordance model (see e.g. \cite{Bahcall2000,deBernardis2000,Tegmark2004}) of a spatially flat  $k=0$ universe composed primarily of dark matter and the dark energy required by the accelerating universe studies of \cite{Riess1998,Perlmutter1999}. 

To  study gravitational fluctuations it is very convenient to use the scalar, vector, tensor expansion of the fluctuation metric that was introduced in  \cite{Lifshitz1946}  and \cite{Bardeen1980} and then widely applied in  perturbative cosmological studies  (see e.g. \cite{Kodama1984,Mukhanov1992,Stewart1990,Ma1995,Bertschinger1996,Zaldarriaga1998,{Bertschinger2006}} and \cite{Dodelson2003,Mukhanov2005,Weinberg2008,Lyth2009,Ellis2012}).  This expansion is based on quantities that transform as 3-dimensional scalars, vectors and tensors, and as such it is particularly well suited to maximally 3-symmetric Robertson-Walker geometries. Even though the scalar, vector, tensor expansion  is defined with respect to three spatial dimensions rather than four spacetime dimensions, nonetheless it leads to fluctuation equations that are composed of combinations of them that are fully 4-dimensionally gauge invariant. They are thus very convenient for cosmological fluctuation theory studies. Thus in this paper we shall follow the gauge invariant procedure that was presented in \cite{Amarasinghe2019,Phelps2019,Mannheim2020,Amarasinghe2021a,Amarasinghe2021b} in which no restriction to a specific choice of gauge was made. (For some other studies that emphasize working with gauge invariant quantities see \cite{Kodama1984,Mukhanov1992,Stewart1990}.) Working throughout solely with gauge invariant quantities we obtain an exact solution  to the standard  model cosmological fluctuation equations in the matter-dominated, hydrodynamic, adiabatic limit. Given this solution  we  show that at low frequencies there are no acoustic oscillations. However, at high frequencies there are. There is thus a maximum wavelength  to these oscillations, with this maximum wavelength providing a horizon for acoustic oscillations.

 \section{Setting up the fluctuation equations}
 \label{S2}
In terms of the convenient conformal time $\eta=\int cdt/a(t)$ where $t$ is the comoving time and $a(t)=\Omega(\eta)$ is the expansion radius,  the background and fluctuation metrics are given by 
\begin{align}
ds^2&=-(g_{\mu\nu}+h_{\mu\nu})dx^{\mu}dx^{\nu}=\Omega^2(\eta)\left[d\eta^2-\frac{dr^2}{(1-kr^2)}-r^2d\theta^2-r^2\sin^2\theta d\phi^2\right]
\nonumber\\
&+\Omega^2(\eta)\left[2\phi d\eta^2 -2(\tilde{\nabla}_i B +B_i)d\eta dx^i - [-2\psi\tilde{\gamma}_{ij} +2\tilde{\nabla}_i\tilde{\nabla}_j E + \tilde{\nabla}_i E_j + \tilde{\nabla}_j E_i + 2E_{ij}]dx^i dx^j\right],
\label{2.1}
\end{align}
as associated with a background geometry with 3-curvature $k$ ($k$ kept here only for generality), 3-metric $\tilde{\gamma}_{ij}$, 3-derivative operator $\tilde{\nabla}_i$, and scalar, vector and tensor metric fluctuations.
In the standard  cosmology one considers a background Einstein tensor $G_{\mu\nu}$ and a background perfect fluid matter energy-momentum tensor $T_{\mu\nu}$ together with fluctuations around them. The matter sources consist of baryons (energy density $\rho_B$), photons ($\rho_{\gamma}$), neutrinos ($\rho_{\nu}$), dark matter ($\rho_D$) and a cosmological constant $\Lambda$.  With the standard model $\Lambda$ contribution only being relevant at current temperatures, at higher redshift we only need to consider the four matter sources, which we label with a suffix $a$ that ranges from 1 to 4.  We take the baryons and dark matter to be nonrelativistic  ($p_B=p_D=0$), and the photons and neutrinos to be relativistic ($p_{\gamma}=\rho_{\gamma}/3$, $p_{\nu}=\rho_{\nu}/3$), with each $\rho$, $p$ pair obeying $\dot{\rho}=-3\dot{\Omega}\Omega^{-1}(\rho+p)$. Since $\rho_{\gamma}$ and $\rho_{\nu}$ both behave as $T^4$ at all temperatures we can set $\rho_{\nu}=N\rho_{\gamma}$ where $N=3(7/8)=21/8$. Similarly, since nonrelativistic $\rho_{B}$ and $\rho_{D}$ both behave as $T^3$ we can set $\rho_{D}=M\rho_{B}$, where in the standard cosmology $M\approx 5$.

In terms of the convenient background and fluctuation quantities 
\begin{align}
R=\frac{8\pi G}{c^4}\sum_{a=1}^{a=4}\rho_a,\qquad P=\frac{8\pi G}{c^4}\sum_{a=1}^{a=4}p_a,\qquad \delta R=\frac{8\pi G}{c^4}\sum_{a=1}^{a=4}\delta \rho_a,\qquad\delta P=\frac{8\pi G}{c^4}\sum_{a=1}^{a=4}\delta p_a,
\label{2.2}
\end{align}
the background and fluctuation equations are $\Delta^{(0)}_{\mu\nu}=0$ and $\Delta_{\mu\nu}=0$ where
\begin{align}
&\Delta^{(0)}_{\mu\nu}=G_{\mu\nu}+(R+P)U_{\mu}U_{\nu}+Pg_{\mu\nu},
\nonumber\\
&\Delta_{\mu\nu}=\delta G_{\mu\nu}+(\delta R+\delta P)U_{\mu}U_{\nu}+\delta Pg_{\mu\nu}+Ph_{\mu\nu}+\frac{8\pi G}{c^4}\sum_{a=1}^{a=4}(\rho_a+p_a)(U_{\mu}\delta U^a_{\nu}+U_{\nu}\delta U^a_{\mu}).
\label{2.3}
\end{align}
With the dot denoting the derivative with respect to $\eta$, the background evolution equations are of the form
 \begin{align}
 &k+\dot{\Omega}^2\Omega^{-2}=\tfrac{1}{3}\Omega^2 R,~~ k-\dot{\Omega}^2\Omega^{-2}+2\ddot{\Omega}\Omega^{-1}=-\Omega^2P,
 ~~ k+2\dot{\Omega}^2\Omega^{-2}-\ddot{\Omega}\Omega^{-1}=\tfrac{1}{2}\Omega^2(R+P).
 \label{2.4}
 \end{align}
If we set $8\pi G\rho_B/3c^4=b/\Omega^3$, $8 \pi G \rho_{\gamma}/3c^4=a/\Omega^4$, these background equations have an exact  $k=0$, $\Lambda=0$ solution
\begin{align}
\Omega(\eta)=((N+1)a)^{1/2}\eta+\tfrac{1}{4}(M+1)b\eta^2.
\label{2.5}
\end{align}
Since $ct=\int d\eta\Omega(\eta)=((N+1)a)^{1/2}\eta^2/2+(M+1)b\eta^3/12$, comoving time and conformal time increase together.

In a 3-momentum mode with separation constant $-q^2$ the key scalar sector components of $\Delta_{\mu\nu}=0$ are (see e.g.  \cite{Phelps2019}):
\begin{align}
& 
\delta \hat{R} \Omega^4= -6 \dot{\Omega}^2 (\alpha-\dot\gamma) + 2 \dot{\Omega} \Omega q^2\gamma,
\label{2.6}
 \end{align}
 \begin{align}
 &\Omega^4 \delta \hat{P}=
 - 2 \dot{\Omega}^2 (\alpha-\dot\gamma)
+2  \dot{\Omega} \Omega(\dot\alpha -\ddot\gamma)+4\ddot\Omega\Omega(\alpha-\dot\gamma)-q^2(\alpha +2\dot\Omega \Omega^{-1}\gamma), 
\label{2.7}
 \end{align}
 \begin{align}
 &\alpha +2\dot\Omega \Omega^{-1}\gamma=0,
\label{2.8}
 \end{align}
 \begin{align}
&
2 \dot{\Omega} \Omega^{-1} (\alpha - \dot\gamma) - 2 k\gamma 
+\frac{8\pi G}{c^4}\Omega^2\sum_{a=1}^{a=4}  (\rho_a+p_a)\hat{X}_a=0,
\label{2.9}
 \end{align}
where $\Omega \tilde{\nabla}_iX_a$ is the longitudinal component of $\delta U^a_i$, and where $\alpha$, $\gamma$ and the hatted quantities  defined by  
\begin{align}
\alpha  &= \phi + \psi + \dot B - \ddot E,\quad \gamma = - \dot\Omega^{-1}\Omega \psi + B - \dot E,\quad \hat{X}_a = X_a-\Omega \dot \Omega^{-1}\psi,
 \nonumber\\
 \delta \hat{\rho}_a&=\delta\rho_a-3(\rho_a+p_a)\psi, \quad \delta \hat{p}_a=\delta p_a+ \dot{\Omega}^{-1}\dot{p}_a\psi \Omega,\quad 
\delta \hat{R}=\delta R-3(R+P)\psi,\quad \delta \hat{P}=\delta P +\dot{\Omega}^{-1}\dot{P}\psi \Omega,
\label{2.10}
\end{align}
are all gauge invariant (see e.g. \cite{Phelps2019,Amarasinghe2021a}). 

To appreciate the structure exhibited in (\ref{2.10}) we note that in (\ref{2.1}) there are four scalar sector metric fluctuations, $\phi$, $\psi$, $B$ and $E$. While there are four gauge transformations of the form $x^{\mu}\rightarrow x^{\mu}+\epsilon^{\mu}$, only two of them affect the scalar sector, viz. $\epsilon_0=-\Omega^2 T$, and the $\Omega^2\tilde{\nabla}_iL$  longitudinal component of $\epsilon_i$, leading to $\phi \rightarrow \phi -\dot{T}-\dot{\Omega}\Omega^{-1}T$, $\psi \rightarrow \psi+\dot{\Omega}\Omega^{-1}T$, $B\rightarrow B+T-\dot{L}$, $E \rightarrow E-L$, $B-\dot{E}\rightarrow B-\dot{E}+T$. Thus out of the four  metric fluctuations there can only be two gauge invariant combinations. As identified by Bardeen \cite{Bardeen1980} they are $\alpha$ and $\gamma$ as given in (\ref{2.10}). Since there are two scalar sector gauge transformations we can use them to set $B=0$ and $E=0$, to then be in the conformal Newton gauge in which $-\dot{\Omega}\Omega^{-1}\gamma=\psi$, and $\alpha=\phi+\psi$. If in addition we impose (\ref{2.8}) we obtain $\phi=\psi$. 

In the perturbed matter sector there are four scalar matter fluctuations, $\delta \rho_a$, $\delta p_a$, $\delta U^a_0$ and $X_a$ for each of the four $a$. Since $g^{\mu\nu}U^a_{\mu}U^a_{\nu}=-1$ for each $a$, it follows that $\delta [g^{\mu\nu}U^a_{\mu}U^a_{\nu}]=0$, from which, without yet choosing a gauge, we  obtain $\delta U^a_0=-\Omega \phi$. With only two scalar sector gauge transformations, we need one extra piece of information for the other  three matter functions. With $\delta{\rho}_a\rightarrow \delta{\rho}_a-T\dot{\rho}_a$, $\delta{p}_a\rightarrow \delta{p}_a-T\dot{p}_a$, $X_a\rightarrow X_a+T$, this is provided by $\psi$ as it transforms into $\psi+\dot{\Omega}\Omega^{-1}T$,  to thereby lead to the gauge invariant $\delta \hat{\rho}_a$, $\delta \hat{p}_a$ and $\hat{X}_a$ as shown in (\ref{2.10}). Thus even without choosing a gauge in (\ref{2.10}) we only need $\phi$ and $\psi$ in order to characterize the scalar sector matter fluctuations. 

Now neither the perturbed Einstein tensor nor the perturbed energy-momentum tensor is separately gauge invariant. Moreover, in and of itself the perturbed $\Delta_{\mu\nu}$ combination given in (\ref{2.3}) is not gauge invariant either, but becomes so when we impose the background  $\Delta^{(0)}_{\mu\nu}=0$. Then $\Delta_{\mu\nu}$ is expressable entirely in terms of gauge invariant quantities, to thus lead to the gauge invariant (\ref{2.6}), (\ref{2.7}), (\ref{2.8}) and (\ref{2.9}).  While we could work in a gauge such as the conformal Newton gauge, we have found it more convenient to work with gauge invariant quantities in the following.

In  the adiabatic perturbations that we study here  the change in each $\rho_a$ is caused solely by the change in the temperature $T$ due to cosmic expansion, so that for adiabatic fluctuations we have
\begin{align}
\frac{\delta \rho_{\gamma}}{\rho_{\gamma}}=\frac{\delta \rho_{\nu}}{\rho_{\nu}}=\frac{4\delta T}{T},\qquad \frac{\delta \rho_{B}}{\rho_{B}}=\frac{\delta \rho_{D}}{\rho_{D}}=\frac{3\delta T}{T}=\frac{3}{4}\frac{\delta \rho_{\gamma}}{\rho_{\gamma}}=\frac{3}{4}\frac{\delta \rho_{\nu}}{\rho_{\nu}}.
\label{2.11}
\end{align}
These relations are general for any epoch that has the indicated background $T^4$ or $T^3$ temperature behavior. Thus in the matter-dominated era in which $\rho_B\gg\rho_{\gamma}$, $\delta\rho_B\gg\delta\rho_{\gamma}$, nonetheless the fractional changes $\delta \rho_{\gamma}/\rho_{\gamma}$ and $\delta\rho_B/\rho_B$ are of the same order of magnitude, and as we shall see, both are relevant  in the  fluctuation equations. Thus even when matter dominates over radiation in the background, it does not do so in the fluctuations.

If for any $p_a$ and $\rho_a$ we have $p_a=w_a\rho_a$, then $\delta p_a=w_a\delta \rho_a$. However, as given in (\ref{2.10}), the relation between $\delta p_a$ and  the gauge invariant $\delta \hat{p}_a$ involves $\dot{p}_a$, and thus in principle involves derivatives of $w_a$. Taking each $w_a$ to be time independent then enables us to set $\delta \hat{p}_a=w_a\delta\hat{\rho}_a$, to thus give
\begin{align}
\frac{\delta \hat{p}_{\gamma}}{\delta \hat{\rho}_{\gamma}}=\frac{1}{3} =\frac{p_{\gamma}}{\rho_{\gamma}},\qquad \frac{\delta \hat{p}_{\nu}}{\delta \hat{\rho}_{\nu}}=\frac{1}{3} =\frac{p_{\nu}}{\rho_{\nu}},\qquad \delta \hat{p}_{B}=\delta \hat{p}_{D}=0.
\label{2.12}
\end{align}

\section{Decoupling the fluctuation equations}
\label{S3}

As it stands, in total there are five scalar sector variables, $\delta\hat{R}$, $\delta\hat{P}$, $\sum_{a=1}^{a=4}  (\rho_a+p_a)\hat{X}_a$, $\alpha$ and $\gamma$, but there are only the four gravitational equations in the scalar sector as given above.  
With $\bar{R}=3\rho_B/4\rho_{\gamma}=3b\Omega/4a$, and with $F_B$ denoting $\delta \hat{\rho}_B/\rho_B=\delta \hat{\rho}_D/\rho_D=3\delta \hat{\rho}_{\gamma}/4\rho_{\gamma}=3\delta \hat{\rho}_{\nu}/4\rho_{\nu}$  as per (\ref{2.11}),  then with (\ref{2.12})  we have a specification of $\delta\hat{P}/\delta\hat{R}$ according to
\begin{align}
&\delta \hat{\rho}_{\gamma}=F_B\frac{\rho_B}{\bar{R}},\qquad \frac{\delta\hat{\rho}_B}{\delta\hat{\rho}_{\gamma}}=\bar{R},
\qquad \delta\hat{R}=\frac{8\pi G}{c^4}F_B\frac{\rho_B}{\bar{R}}\left(N+1+(M+1)\bar{R}\right),
\nonumber\\
&\delta\hat{P}=\frac{8\pi G}{c^4}F_B\frac{\rho_B}{3\bar{R}}(N+1),\qquad \frac{\delta\hat{P}}{\delta\hat{R}}=\frac{(N+1)}{3(N+1+\bar{R}(M+1))},
\label{3.1}
\end{align}
with the system then becoming solvable. Of the gravitational fluctuation equations, it is striking that only one, viz. (\ref{2.9}), involves the $\hat{X}_a$. Thus very conveniently we can treat the other three equations without any reference to the $\hat{X}_a$ at all. While these other three equations contain no explicit reference to $k$, they still have an implicit dependence on $k$ as it appears in the background  (\ref{2.4}). Thus whenever the contribution of $k$ to the background evolution equations is negligible the solutions to all of the fluctuation equations except the $\hat{X}_a$ will be the same as if $k$ actually is zero.  
 
 Given (\ref{3.1}), (\ref{2.6}), (\ref{2.7}) and (\ref{2.8}) we now obtain
\begin{align}
3(N+1+(M+1)\bar{R})(- 2 \dot{\Omega}^2 (\alpha-\dot\gamma)
+2  \dot{\Omega} \Omega(\dot\alpha -\ddot\gamma)+4\ddot\Omega\Omega(\alpha-\dot\gamma))=(N+1)( -6 \dot{\Omega}^2 (\alpha-\dot\gamma) + 2 \dot{\Omega} \Omega q^2\gamma).
\label{3.2}
\end{align}
Finally, again through use of (\ref{2.8}) we obtain
\begin{align}
&\gamma\left[(N+1)(6\dot{\Omega}^3 -18\ddot{\Omega}\dot{\Omega}\Omega-\dot{\Omega} \Omega^2 q^2)+(M+1)\bar{R}(12\dot{\Omega}^3 -18\ddot{\Omega}\dot{\Omega}\Omega)\right]
\nonumber\\
&-\dot{\gamma}\left[(N+1)(6\dot{\Omega}^2\Omega+6\ddot{\Omega}\Omega^2)+(M+1)\bar{R}(3\dot{\Omega}^2\Omega+6\ddot{\Omega}\Omega^2)\right]
-3\dot{\Omega}\Omega^2\ddot{\gamma}\left[N+1+(M+1)\bar{R}\right].
 \label{3.3}
\end{align}
With (\ref{3.3}) we have decoupled the fluctuation equations, and we note that (\ref{3.3}) is  exact without approximation in any era (radiation or matter), no matter what the form for the background $\Omega$ (not only with a possible $k$ but also even with $\Lambda$ since a constant $\Lambda$ has no fluctuation). We now  solve (\ref{3.3}) analytically in the matter-dominated era.

\section{Exact solution}
\label{S4}
In the matter-dominated era we have $\Omega=(M+1)b\eta^2/4$ and $\bar{R}=3\rho_B/4\rho_{\gamma}=(3b/4a)\Omega=d\eta^2 \gg 3/4$, where $d=3b^2(M+1)/16a$.  We shall solve the theory exactly in the matter-dominated era, and note the relevance of the solution since the dominant contribution to the measured temperature anisotropy in the CMB is coming from last scattering, with last scattering occuring after  matter-radiation equality. Moreover,  since the full matter plus radiation $\Omega$ is given by $\Omega=[b(M+1)\eta^2/4]\big{[}1+4a^{1/2}(N+1)^{1/2}/b(M+1)\eta\big{]}$, we could solve for $\Omega=b(M+1)\eta^2/4$ and then perturb around it via the $4a^{1/2}(N+1)^{1/2}/b(M+1)\eta$ term. (While we have found an exact solution in the matter-dominated era we have not found one in the radiation-dominated era.)

With $\Omega=b(M+1)\eta^2/4$  (\ref{3.3}) takes the form
\begin{align}
&\gamma\left[-(N+1)(12+\eta^2 q^2)+12(M+1)d\eta^2\right]
+\dot{\gamma}\left[-18(N+1)\eta-12\eta(M+1)d\eta^2\right]
\nonumber\\
&+\ddot{\gamma}\left[-3(N+1)\eta^2 -3\eta^2(M+1)d\eta^2\right]
=0.
 \label{4.1}
\end{align}
Quite remarkably, (\ref{4.1})  has an exact solution. With  
\begin{align}
&A=\left(\frac{25}{16}-\frac{(N+1)q^2}{12d(M+1)}\right)^{1/2}
\label{4.2}
\end{align}
we find an exact hypergeometric function solution  of the form
\begin{align}
\gamma(\eta)&=\frac{c_1Y^{-2}}{\eta^4}F(-5/4-A,-5/4+A;-1/2;-Y\eta^2)-\frac{c_2Y^{-1/2}}{\eta}F(1/4-A,1/4+A;5/2;-Y\eta^2)
\label{4.3},
\end{align}
where $Y=d(M+1)/(N+1)$, $Y\eta^2=\bar{R}(M+1)/(N+1)$, and $c_1$ and $c_2$ are constants. 

Now the hypergeometric functions obey \cite{digital}
\begin{align}
\frac{d}{dz}F(a,b;c;z)=\frac{ab}{c}F(a+1,b+1;c+1;z),
\label{4.4}
\end{align}
Thus for $\dot{\gamma}$ and $\ddot{\gamma}$ we obtain
\begin{align}
&\dot{\gamma}=-\frac{4c_1Y^{-2}}{\eta^5}F(-5/4-A,-5/4+A;-1/2;-Y\eta^2)
-\frac{c_1Y^{-1}(16A^2-25)}{4\eta^3}F(-1/4-A,-1/4+A;1/2;-Y\eta^2)
\nonumber\\
&+\frac{c_2Y^{-1/2}}{\eta^2}F(1/4-A,1/4+A;5/2;-Y\eta^2)
+\frac{c_2Y^{1/2}(1-16A^2)}{20}F(5/4-A,5/4+A;7/2;-Y\eta^2),
\label{4.5}
\end{align}
\begin{align}
&\ddot{\gamma}=\frac{20c_1Y^{-2}}{\eta^6}F(-5/4-A,-5/4+A;-1/2;-Y\eta^2)
+\frac{7c_1Y^{-1}(16A^2-25)}{4\eta^4}F(-1/4-A,-1/4+A;1/2;-Y\eta^2)
\nonumber\\
&+\frac{c_1(16A^2-25)(1-16A^2)}{16\eta^2}F(3/4-A,3/4+A;3/2;-Y\eta^2)
\nonumber\\
&-\frac{2c_2Y^{-1/2}}{\eta^3}F(1/4-A,1/4+A;5/2;-Y\eta^2)
-\frac{c_2Y^{1/2}(1-16A^2)}{20\eta}F(5/4-A,5/4+A;7/2;-Y\eta^2)
\nonumber\\
&-\frac{c_2Y^{3/2}\eta (1-16A^2)(25-16A^2)}{560}F(9/4-A,9/4+A;9/2;-Y\eta^2).
\label{4.6}
\end{align}
In addition the hypergeometric function obeys 
the contiguous hypergeometric function relation \cite{digital}
\begin{align}
&z(1-z)(a+1)(b+1)F(a+2,b+2;c+2;z)
+(c-(a+b+1)z)(c+1)F(a+1,b+1;c+1;z)
\nonumber\\
&-c(c+1)F(a,b;c;z)=0.
\label{4.7}
\end{align}
From (\ref{4.7}) we obtain 
\begin{align}
&-Y\eta^2(1+Y\eta^2)\left(\frac{1}{16}-A^2\right)F(3/4-A,3/4+A;3/2;-Y\eta^2)
\nonumber\\
&-\frac{1}{4}(1+3Y\eta^2)F(-1/4-A,-1/4+A;1/2;-Y\eta^2)+\frac{1}{4}F(-5/4-A,-5/4+A;-1/2;-Y\eta^2)=0,
\nonumber\\
&-Y\eta^2(1+Y\eta^2)\left(\frac{25}{16}-A^2\right)F(9/4-A,9/4+A;9/2;-Y\eta^2)
\nonumber\\
&+\frac{7}{4}(5+3Y\eta^2)F(5/4-A,5/4+A;7/2;-Y\eta^2)-\frac{35}{4}F(1/4-A,1/4+A;5/2;-Y\eta^2)=0,
\label{4.8}
\end{align}
and with these relations we can readily check that the form for $\gamma$ as given in (\ref{4.3}) does indeed obey (\ref{4.1}). 

We note that is only through the dependence of (\ref{4.1}) on $q^2$ that the solution is nontrivial, since $\gamma(q^2=0)=c_1Y^{-2}\eta^{-4}-c_2Y^{-1/2}(\eta^{-1}+3Y\eta/5)$ is just power behaved. Similarly if set $N+1$ to zero (i.e., no photons or neutrinos at all), then from (\ref{5.1}) below with $Y \rightarrow \infty$ we obtain the power-behaved $\gamma \sim \eta+\eta^{-4}$. Thus without photons or neutrinos there  can be no baryon acoustic oscillations. Moreover, in (\ref{4.1}) even while $(M+1)d\eta^2 \gg (N+1)$ in the matter-dominated era, it will not be the case that $(M+1)d\eta^2 \gg (N+1)q^2\eta^2$ if $q^2$ is large enough, with $q^2$ then being relevant and $\delta \hat{P}$ then not being small in the matter-dominated era even as the background $P$ is then negligible.

From the equations for $\delta\hat{R}$ given in (\ref{3.1}) and (\ref{2.6}), and with $\Omega(\eta)=(M+1)b\eta^2/4$,  we have 
\begin{align}
&[N+1+(M+1)\bar{R}]F_B= \frac{\left[32ad(M+1)\eta\gamma +8ad(M+1)\eta^2\dot{\gamma}+(4ad/3)(M+1)\eta^3q^2\gamma\right]}{4a}
\nonumber\\
&=\frac{c_1(N+1)^2q^2}{3(M+1)d\eta}\bigg{(}F(-5/4-A,-5/4+A;-1/2;-Y\eta^2)
+\frac{2}{(M+1)^2}F(-1/4-A,-1/4+A;1/2;-Y\eta^2)\bigg{)}
\nonumber\\
&-\frac{c_2d^{1/2}(N+1)^{1/2}(M+1)^{1/2}}{3}\bigg{(}(18+\eta^2q^2)F(1/4-A,1/4+A;5/2;-Y\eta^2)
\nonumber\\
&+\frac{2\eta^2}{5(N+1)}\left(18d(M+1)-(N+1)q^2\right)F(5/4-A,5/4+A;7/2;-Y\eta^2)\bigg{)}.
\label{4.9}
\end{align}
From (\ref{2.9}) we can only determine the combination 
\begin{align}
\sum_{a=1}^{a=4}(\rho_a+p_a)\hat{X}_a=\frac{4}{3}\rho_{\gamma}\hat{X}_{\gamma}+\rho_B\hat{X}_B+\frac{4N}{3}\rho_{\gamma}\hat{X}_{\nu}+M\rho_B\hat{X}_D
=\frac{8c^4(4\gamma+\eta\dot{\gamma})}{\pi Gb^2(M+1)^2\eta^4},
\label{4.10}
\end{align}
where $\gamma$ and $\dot{\gamma}$ are respectively given in (\ref{4.3}) and (\ref{4.5}) and where we have set $k=0$ in (\ref{2.9}). The relations (\ref{4.3}), (\ref{4.9}) and (\ref{4.10}) are exact in a hydrodynamic, adiabatic,   matter-dominated era and constitute our main result. 

To determine the individual $\hat{X}_a$  requires additional dynamical information beyond that contained in the gravitational fluctuation equations themselves. This information would have to  come from the individual $\delta T^{\mu\nu}_a$  of each of the four matter field species and is discussed in the extended version of this paper that is given in \cite{Mannheim2024}. Specifically, it was noted there that even though each $\delta T^{\mu\nu}_a$ is not gauge invariant, the scalar sector of each $\delta[\nabla_{\mu}T^{\mu\nu}_a]$ is, viz. 
 \begin{align}
&c \delta[\nabla_{\mu}T_a^{\mu 0}]=\Omega^{-2}\left[\nabla_0\delta\hat{\rho}_a+3\dot{\Omega}\Omega^{-1}(\delta \hat{\rho}_a+\delta \hat{p}_a)-(\rho_a+p_a)q^2(\hat{X}_a-\gamma)\right],
 \nonumber\\
 &c \delta[\nabla_{\mu}T_a^{\mu i}]=\Omega^{-2}\tilde{\gamma}^{ij}\tilde{\nabla}_j[\left[\delta\hat{p}_a+\dot{p}_a\hat{X}_a+(\rho_a+p_a)[\alpha-\dot{\gamma}+\dot{\hat{X}}_a+\dot{\Omega}\Omega^{-1}\hat{X}_a]\right].
 \label{4.11}
 \end{align}
If we set any of these individual covariant derivatives to zero, we can solve for individual $\hat{X}_a$. As discussed in \cite{Mannheim2024}, this task is quite straightforward once we already know $\gamma$ and $\alpha$ (for instance $\delta[\nabla_{\mu}T_a^{\mu 0}]=0$ entails that  $\hat{X}_a=\dot{F}_B/q^2+\gamma$), though it would be much more difficult if we were to start with the vanishing of  individual $\delta[\nabla_{\mu}T_a^{\mu \nu}]$ or  combinations thereof, and then tried to obtain $\gamma$ and $\alpha$ from them.

\section{Minimum Frequency  for Acoustic Oscillations}
\label{S5a}

Now that we have the exact solution we can analyze the acoustic oscillations that are thought to be associated with the baryon-photon plasma, and we find that  such oscillations can only occur at high frequencies. Specifically, as a function of $z$ the $F(a,b;c;z)$ hypergeometric functions have no or possibly only a few isolated zeroes on the negative $z$ axis if the $a$, $b$ and $c$ associated with (\ref{4.3}) and (\ref{4.9}) are all real \cite{digital}. (Maxima and minima in a given $F(a,b;c;z)$ are located at the zeroes of its derivative as given in (\ref{4.4}), i.e., by the zeroes of $F(a+1,b+1;c+1;z)$.) However, the $F(a,b;c;z)$ of interest to us here do have zeroes on the negative $z=-Y\eta^2$ axis if $A$ as given in (\ref{4.2}) is imaginary, something which occurs if $q^2/d>75(M+1)/4(N+1) \approx 30$, with there thus being a   minimum frequency $q_{MIN}=[75d(M+1)/4(N+1)]^{1/2}=15b(M+1)/8[(N+1)a]^{1/2}$ for acoustic oscillations. 

To characterize these oscillations we note that hypergeometric functions have a leading asymptotic behavior \cite{digital}
\begin{align}
&\frac{ \sin(\pi(b-a))}{\pi}F(a,b;c;z)=\frac{(-z)^{-a}}{\Gamma(b)\Gamma(c-a)}F(a,a-c+1;a-b+1;1/z)-\frac{(-z)^{-b}}{\Gamma(a)\Gamma(c-b)}F(b,b-c+1;b-a+1;1/z)
\nonumber\\
&
\rightarrow \frac{(-z)^{-a}}{\Gamma(b)\Gamma(c-a)}\left(1+\frac{a(a-c+1)}{(a-b+1)z}\right)-\frac{(-z)^{-b}}{\Gamma(a)\Gamma(c-b)}\left(1+\frac{b(b-c+1)}{(b-a+1)z}\right)
\label{5.1}
\end{align}
as $z\rightarrow\infty$. In consequence, at large $Y^{1/2}\eta$ the $c_1$ and $c_2$ terms in $\gamma$ both have the same $[Y^{1/2}\eta]^{2A-3/2}$ power behavior in conformal time, and thus behave as  $t^{2A/3-1/2}$ in comoving time.  At large $q^2/d$ we can set set $A=iB$ where $B=q/2(3Y)^{1/2}$ is real. Then the solutions oscillate as $e^{iB\log(Y\eta^2)}$ or $e^{(2iB/3)\log t}$.  Moreover, it is explicitly this $e^{iB\log(Y\eta^2)}$ behavior that is to be expected for acoustic oscillations. Specifically, we note that with $\bar{R}=d\eta^2$ we recognize $Y\eta^2$ as $\bar{R}(M+1)/(N+1)$, with this factor being none other than the factor that appears in $\delta\hat{P}/\delta \hat{R}$ in (\ref{3.1}), so that we can set $\delta\hat{P}/\delta \hat{R}=1/(3(1+Y\eta^2))$. In the matter-dominated large $\bar{R}$
limit we can thus set $\delta\hat{P}/\delta \hat{R}=1/3Y\eta^2$. Thus  if we identify $\delta\hat{P}/\delta \hat{R}=v^2/c^2$, where $v$ is the sound velocity, then we can identify $v^2/c^2$ as the time-dependent $1/3Y\eta^2$. Acoustic oscillations then oscillate according to 
\begin{align}
\exp\left[i\int \frac{vqd\eta}{c}\right]=\exp\left[i\int\frac{ q d\eta }{(3Y)^{1/2}\eta}\right]=\exp\left[\frac{iq \log \eta }{(3Y)^{1/2}}\right]=\exp\left[\frac{4iq a^{1/2}(N+1)^{1/2} \log \eta }{3b(M+1)}\right].
\label{5.2}
\end{align}
While the validity of (\ref{5.2}) only requires that  $(N+1)q^2/12d(M+1)>25/16$, we can recover (\ref{5.1}), since  when 
 $q^2/d$ is large, (\ref{5.2}) is of the form $e^{2iB\log \eta}$, just as required of (\ref{5.1}).  The matter-dominated era approach presented here allows us to analytically monitor lower frequency departures from the high frequency WKB, since for frequencies $q$ greater than $q_{MIN}$ but not asymptotically greater than $q_{MIN}$  there will already be oscillations. (Acoustic oscillations were also found in \cite{Bertschinger2006} using an asymptotic  high frequency WKB approximation and matter-radiation equality.)
 
Since every hypergeometric function in both $\gamma(\eta)$ as given in (\ref{4.3}) and in $F_B$ as given in (\ref{4.9}) have both the same $z= -Y\eta^2$ and the same $A$, each and every one of these functions oscillate this same way at large $q^2$. Since these oscillations do occur in $F_B$ they will appear in $\delta \rho_B/\rho_B$ and not just in the metric coefficients. The analysis that we have given here supports the identification of $\delta\hat{P}/\delta \hat{R}$ with the square of the sound velocity in curved space cosmological perturbation theory, just as it holds in flat space kinetic theory. However, our analysis only leads to acoustic oscillations at high frequency ($q^2/d>30$), but not at low frequency ($q^2/d<30$). 

With there being a minimum $q_{MIN}$ for acoustic oscillations and a thus associated  maximum wavelength $\lambda_{MAX}=2\pi/q_{MIN}$, we should consider whichever is the smaller of  this $\eta$-independent $\lambda_{MAX}$ or the standard $\eta$-dependent but $q$-independent $\Omega(\eta)\int_0^\eta d\eta/(3(1+\bar{R}))^{1/2}=\Omega(\eta)\int_0^\eta d\eta v/c$ to be  the sound horizon.

Since there are acoustic oscillations at all, these oscillations break the smoothness aspect of the adiabatic approximation as they lead to oscillating rather than smooth temperature fluctuations, with the parameter $A$ as defined in (\ref{4.2}) having a square root branch point at $q^2=q^2_{MIN}$ in the complex $q^2$ plane \cite{footnoteA}. Thus for frequencies above $q_{MIN}$ we have to take nonhydrodynamic, i.e.,  viscous, effects into consideration. Below this minimum frequency, i.e., for wavelengths greater that the maximum wavelength $\lambda_{MAX}=2\pi/q_{MIN}$ of these oscillations, we can continue to use the hydrodynamic approximation (just as it holds in flat space kinetic theory at long wavelengths). As long as particles are within the causal horizon, particles that are separated by more than  $\lambda_{MAX}$ will be able to cause the collision integral in the Boltzmann equation  that describes the statistics of the system to vanish by having particles enter and exit every region of phase space at exactly the same rate, which thereby keeps the system in hydrodynamic equilibrium. For distances shorter than $\lambda_{MAX}$ the particles will not be able to make the collision integral vanish and there will be nonhydrodynamic viscous effects \cite{footnoteB}. 

We recall that in standard flat space kinetic theory the hydrodynamic approximation is only a low frequency, long wavelength limit (to get all of a gas to thermalize we need long wavelengths so as to encompass the entire gas), with viscous effects modifying the hydrodynamic limit at large frequency. However in the analog cosmological case, as we go to higher $q^2$, there is actually a self-quenching effect since (see e.g. \cite{Weinberg2008}) viscosity leads to Silk damping \cite{Silk1972}, which together with Landau damping, leads to exponential ($\sim e^{-q^2}$) suppression of the large $q^2$ modes. This then enables a hydrodynamic treatment of the temperature anisotropy in the CMB at all $q^2$ to potentially be viable after all, though even so there will still only be acoustic oscillations at high frequency. The concerns expressed here are also relevant to studies of the low frequency domain of the matter fluctuation power spectrum, both in general, as well as possibly having a connection to the nonrelativistic approach described in \cite{Ivanov2022}.

\section{A curvature-dominated exact solution}
\label{S9}

Another case where we have been able to find an exact solution is the 3-curvature-dominated case. In this case we take $k$ to dominate in the background (\ref{2.4}), so that we have to solve $k+\dot{\Omega}^2/\Omega^2=0$. This then requires that $k$ be negative, so we set it equal to $-1/L^2$, with the solution then being $\Omega(\eta)=e^{\eta/L}$, so that $a(t)=ct/L$, $H(t)=1/t$, $\Omega(\eta)=a(t)=c/LH(t)=e^{\eta/L}$. With $k<0$ the spatial sector separation constant is given by $q^2=(\tau^2+1)/L^2$ where $\tau \in (0,\infty)$ \cite{Mannheim2020}, with a negative value for $k$ being favored in the conformal gravity theory described in  \cite{Mannheim2006}, though in the conformal case there is no need for dark matter. In the $\Omega=e^{\eta/L}$ standard gravity solution $\bar{R}$ takes the value $3be^{\eta/L}/4a$. With this form for $\Omega$ and $\bar{R}$  (\ref{3.3}) takes the form
\begin{align}
&\gamma[(N+1)(12+L^2q^2)+18b(M+1)e^{\eta/L}/4a]+L\dot{\gamma}[12(N+1)+27b(M+1)e^{\eta/L}/4a]
\nonumber\\
&+L^2\ddot{\gamma}[3(N+1)+9b(M+1)e^{\eta/L}/4a]=0,
\label{6.1}
\end{align}
and has an exact hypergeometric solution
\begin{align}
\gamma(\eta)&=d_1e^{-2\eta/L}e^{-iq\eta/\sqrt{3}} F(-1-iqL/\sqrt{3};-iqL/\sqrt{3};1-2iqL/\sqrt{3};-e^{\eta/L}3b(M+1)/(4a(N+1)))
\nonumber\\
&+d_2e^{-2\eta/L}e^{iq\eta/\sqrt{3}} F(-1+iqL/\sqrt{3};+iqL/\sqrt{3};1+2iqL/\sqrt{3};-e^{\eta/L}3b(M+1)/(4a(N+1))),
\label{6.2}
\end{align}
where $d_1$ and $d_2$ are constants. With $\Omega=e^{\eta/L}$  the factor $e^{\eta/L}3b(M+1)/(4a(N+1))$ equals  $\bar{R}(M+1)/(N+1)$. Thus just like the matter-dominated (\ref{4.3}),  this factor is none other than the factor that appears in  $\delta \hat{P}/\delta \hat{R}$ as given in (\ref{3.1}). 

The solution in (\ref{6.2}) possesses two sources for oscillations. First there are  oscillations due to the overall factor $e^{\mp iq\eta/\sqrt{3}}=e^{\mp i(qL/\sqrt{3})\log (ct/L)}$, i.e., oscillations that are associated with massless particles even though massive matter fields contribute to (\ref{6.1}). The second source is from the hypergeometric functions themselves. Since $iqL/\sqrt{3}$ is imaginary for any value of $q$, unlike the matter-dominated case discussed in Sec. \ref{S4}, the hypergeometric functions in (\ref{6.2}) will have zeroes on the $-e^{\eta/L}3b(M+1)/(4a(N+1))$ axis even at low frequency, (i.e., unlike the behavior of the $A$ parameter given in (\ref{4.2}), this time there is no complex $q^2$ plane branch point). However, according to (\ref{5.1}) and in analog to the matter-dominated case, there will also be high frequency oscillations, which in this case are of the form $e^{\pm iq\eta/\sqrt{3}}$. These oscillations will exactly cancel those due to the overall $e^{\mp iq \eta/\sqrt{3}}$ factor in (\ref{6.2}), to then leave this overall oscillation factor and lower frequency oscillations due to the zeroes of the hypergeometric functions to dominate. We note that the overall $e^{ i(qL/\sqrt{3})\log (ct/L)}$ factor is just like the $e^{(2iB/3)\log t}$ factor found for matter-dominated fluctuations in Sec. \ref{S4}, even being equal to it up to a phase  if $L=1/3Y^{1/2}$.

In both (\ref{6.1}) and (\ref{6.2})  we have not switched off fluctuations in the photon, neutrino, baryon and dark matter fields, as the presence of the $3b(M+1)/4a(N+1)$ factor shows. Rather, we have taken the 3-curvature $k$ to dominate in the background evolution equations. In fact, the 3-curvature $k$ cannot dominate in the fluctuation equations, since, as noted above, the derivation of our key (\ref{3.3}) equation and thus (\ref{6.1}) for $\gamma$ only involved the $k$-independent gravitational fluctuation equations given in (\ref{2.6}), (\ref{2.7}) and (\ref{2.8}).

\section{Summary}
\label{S10}

In the thermal evolution of the universe dark matter and neutrinos do not interact with the baryons and photons to any great degree, so that fluctuations in their energy densities are adiabatic, i.e., they are controlled solely by temperature fluctuations, so that $\delta \rho_D\sim 3T^2\delta T$ and $\delta \rho_{\nu}\sim 4T^3\delta T$. However, because of Thomson scattering baryons and photons do interact with each other and this affects fluctuations in their energy densities. These fluctuations can be described by the Boltzmann equation, with thermal equilibrium eventually being reached at long times. In this thermal equilibrium the distribution  function causes the collision integral on the right-hand side of the Boltzmann equation to vanish (by balancing the flow of particles into and out of any given region of phase space), while causing the total time derivative on the left-hand side of the Boltzmann equation to be describable by a fluctuation-dependent energy-momentum tensor. At this point the baryon and photon energy densities become adiabatic  so that the four dark matter,  neutrino, photon and baryon species are then all in hydrodynamic equilibrium with each other.

In the standard cosmological model we have obtained closed form expressions for the dynamical variables in this hydrodynamic phase and have been able to solve them analytically in the matter-dominated era. The solution we find  is in the form of hypergeometric functions. Such functions do not  oscillate at low frequencies but do oscillate at the sound velocity at high frequencies. At these higher frequencies viscosity effects become significant. However, through Silk and Landau damping they actually self quench as viscosity suppresses high frequencies. Thus in studies of standard model cosmological fluctuations the hydrodynamic limit can be used to a reasonable degree at all frequencies.

\begin{acknowledgments}
We acknowledge very helpful  discussions with T. Liu, D. A. Norman, C. M. Bender, E. Bertschinger and A. Olde Daalhuis.
\end{acknowledgments}


\begin{thebibliography}{99}
\bibitem{footnoteA} The branch point identified here is somewhat unusual. While $A$ itself has a complex $q^2$ plane branch cut that begins at $q^2=q^2_{MIN}$, to thus have a two-sheeted Riemann sheet structure, this not the case for $\gamma$. Specifically, because of the  power series expansion $F(a,b;c;z)=1+ab/1!c+a(a+1)b(b+1)/2!c(c+1)+....$ it follows that the hypergeometric functions present in the solution for $\gamma(\eta)$ are only functions of $A^2$, and unlike $A$, $A^2$ has no branch point. While there are still two Riemann sheets in the complex $q^2$ plane, $\gamma$ takes the same value on both of them. $\gamma$  thus has a branch point but no branch cut even while $A$ itself does have a branch cut. Even though $\gamma$ has what is known as a removable singularity, its two-sheeted Riemann sheet  structure remains, and still indicates a breakdown of the adiabatic approximation.
\bibitem{footnoteB} To get a numerical estimate for $q_{MIN}=15b(M+1)/8[(N+1)a]^{1/2}$, we note that  in comoving time in the matter-dominated era the expansion radius $a(t)$  obeys $(da(t)/dt)^2/c^2=(8\pi G/3c^4)a^2(t)(M+1)\rho_B=b(M+1)/a(t)$, with solution $a(t)=\sigma (t/t_0)^{2/3}$, where $t_0$ is the current time, $\sigma$ is a dimensionless  integration constant that is set by an initial condition, and $b(M+1)=4\sigma^3/9c^2t_0^2$. In this solution the current era Hubble parameter is given by $H_0=2/3t_0$. In consequence $b(M+1)=\sigma^3H_0^2/c^2$. With the current value of $H_0$ taken to be $72~\text{km}~\text{sec}^{-1}~\text{Mpc}^{-1}$ we obtain $b(M+1)=6.05\sigma^3\times 10^{-57}~\text{cm}^{-2}$. With $\rho_B/\rho_{\gamma}=b\sigma(t/t_0)^{2/3}/a$, from its current value of $(M+1)\rho_B/\rho_{\gamma}= 0.3\times  8.76\times 10^{-9}/4.17\times 10^{-13}=6.30\times 10^3$ we obtain $a=0.96\sigma^4\times 10^{-60}~\text{cm}^{-2}$. Hence $q_{MIN}=6.08 \sigma \times 10^{-27}~\text{cm}^{-1}$, so that $\lambda_{MAX}=2\pi/q_{MIN}=1.04\sigma^{-1}\times  10^{27}~\text{cm}$. If we set $\sigma=1$ as is common practice, we recognize $\lambda_{MAX}=1.04\times 10^{27}~\text{cm}=337~\text{Mpc}$ as a cosmological scale.
\end{thebibliography}
\end{document}